\documentclass [12pt] {article}
\usepackage{graphicx}
\usepackage{epsfig}
\def\lsim{\raise0.3ex\hbox{$<$\kern-0.75em\raise-1.1ex\hbox{$\sim$}}}
\def\gsim{\raise0.3ex\hbox{$>$\kern-0.75em\raise-1.1ex\hbox{$\sim$}}}

\newcommand{\be}{\begin{equation}}
\newcommand{\ee}{\end{equation}}
\newcommand{\bd}{\begin{displaymath}}
\newcommand{\ed}{\end{displaymath}}
\newcommand{\ba}{\begin{array}}
\newcommand{\ea}{\end{array}}
\newcommand{\bt}{\begin{tabular}}
\newcommand{\et}{\end{tabular}}
\newcommand{\bc}{\begin{center}}
\newcommand{\ec}{\end{center}}

\begin {document}
\vskip -1.cm
%\hspace{8.cm} Oct 12th, 2009
\vskip 0.5cm
\vskip -0.1cm
\begin{center}
{\bf Production of Strange Secondaries in High Energy $\Sigma^-A$ Collisions}

\vskip 0.5 truecm

G.H. Arakelyan$^*$, A.B. Kaidalov$^{**}$, C. Merino$^{***}$, and 
Yu.M. Shabelski$^{****}$ \\

\vspace{.5cm}
 
$^{*}$ Yerevan Physics Institute \\
Armenia
%Yerevan 0036, Armenia \\
E-mail: argev@mail.yerphi.am

\vspace{.2cm}

$^{**}$ Institute of Theoretical and Experimental Physics \\ 
Moscow, Russia \\
%Moscow 117259, Russia \\
E-mail: kaidalov@itep.ru
\vspace{.2cm}

$^{***}$ Departamento de F\'\i sica de Part\'\i culas and \\
Instituto Galego de F\'\i sica de Altas Enerx\'\i as \\
Universidade de Santiago de Compostela \\
%15782 Santiago de Compostela \\
Galiza, Spain \\
E-mail: merino@fpaxp1.usc.es

\vspace{.2cm}

$^{****}$ Petersburg Nuclear Physics Institute \\
Russia \\
%Gatchina, St.Petersburg 188350, Russia \\
E-mail: shabelsk@thd.pnpi.spb.ru

\end{center}

\vskip 1. truecm

\begin{center}
{\bf Abstract}
\end{center}
We describe the WA89 Collaboration experimental
data on $\Lambda$, $\Sigma^-$, $\Sigma^+$, $\Xi^-$, and $\Omega^-$ baryons,
and $\bar{\Lambda}$ and $\overline{\Xi}^+$ antibaryons production in $\Sigma^-$
collisions with C and Cu targets at 345~GeV/c ($\sqrt{s_{\Sigma N}} \approx 25.5$~GeV) in the frame
of the Quark-Gluon String Model. How the theoretical results compare to the
experimental data is discussed. Finally, some relations among the values of the model parameters
obtained with the help of quark combinatorics are presented.

\vskip 1cm


\vskip 2.5 truecm

\newpage
\section{Introduction}
%\vskip 0.4 truecm

The Quark-Gluon String Model (QGSM) and the Dual Parton Model (DPM)
are based on the Dual Topological Unitarization (DTU) and they
quantitatively describe [1-6] many features of high energy production processes,
including the inclusive spectra of different secondary hadrons, their
multiplicities and multiplicity distributions, etc., both in
hadron-nucleon and hadron-nucleus collisions at fixed target energies.
QGSM and DPM account for the main features of secondary production at collider energies.

In the QGSM, high energy interactions are considered as proceeding via the
exchange of one or several Pomerons, and all elastic and inelastic processes
result from cutting through or between Pomerons~\cite{AGK}. Each cut Pomeron
leads to the production of two strings  of secondaries. The inclusive spectra of
hadrons in the final state of the collision are related to the corresponding 
fragmentation functions of the quarks
and diquarks at the end of the strings. These fragmentation functions are
constructed by using the Reggeon counting rules~\cite{Kai}.

To study the interaction with a nuclear target the Multiple Scattering
Theory (Gribov-Glauber Theory) is used, and, thus, this
interaction is considered as the superposition of the interactions
of the incident hadron with different nucleons in the target~\cite{KTMS,Sh1}.

In previous papers [1-6], where the secondary production by proton, pion,
and kaon beams was reasonably described, the model parameters were already
fixed by comparison of the theoretical calculations with experimental data.

Here we consider the production of secondaries in
$\Sigma^-C$ and $\Sigma^-Cu$ collisions and we present the comparison of the
theoretical results with the experimental data at 345 GeV/c obtained by the
WA89 Collaboration~\cite{WA89}. To include in our description interactions out
of $\Sigma^-$ beams we have to deduce the expressions both of the momentum
distribution function of $ds$ diquark in $\Sigma^-$ hyperon, and of its
fragmentation functions into secondary hadrons.

However, in the case of $\Sigma^-$ beam the agreement with the experimental data
of the calculations obtained by using these standard fragmentation functions is
not good enough. The agreement becomes better when some additional polynomial factors
are included into our fragmentation functions, but it is not clear whether this is a result
of some special structure of the strange baryons and their resonances, or it is simply
connected to possible experimental inconsistencies.

\section{Production of secondaries on nuclear targets in QGSM}

In QCD hadrons are composite bound state configurations built up from the
quark $\psi_i(x), i = 1,...N_c$, and gluon $G^{\mu}_a(x), a = 1,...,N_c^2-1$,
fields. In string models baryons are considered as configurations
consisting of three strings attached to three valence quarks and connected at
one point (small volume) $x$, called the ``string junction" (SJ) [11-13]. The 
corresponding wave function can be written as
\begin{eqnarray}
\vert B \rangle &=& \psi_i(x_1)\cdot\psi_j(x_2)\cdot\psi_k(x_3)\cdot J^{ijk}(x,x_1,x_2,x_3) \;,\nonumber\\ 
J^{ijk}(x,x_1,x_2,x_3) &=& \Phi^i_{i'}(x_1,x) \cdot \Phi_{j'}^j(x_2,x) \cdot
\Phi^k_{k'}(x_3,x) \cdot \epsilon^{i'j'k'} \;.
\end{eqnarray}

\noindent
Here the operator $\Phi^i_{i'}(x_1,x)$ represents the gluon field string with
endpoints at $x_1$ and $x$. Such a ``star" (or Y configuration of the baryon
wave function $\vert B \rangle$ is favoured~\cite{Art,RV} with respect to the
also possible ``triangle" (or \,$\Delta$) configuration. 

Let us discuss in more detail the processes in which one or
several Pomerons are exchanged. Each exchanged Pomeron in hadron-nucleon and
hadron-nucleus interaction corresponds to a cylindrical diagram that,
when cut, produces two showers of secondaries. The inclusive spectrum
of secondaries is determined by the convolution of diquark, valence quark,
and sea quark distributions, $u(x,n)$, in the incident particles with the
fragmentation functions, $G(z)$, of quarks and diquarks into the different
hadrons. The diquark and quark distribution functions depend on the number of cut
Pomerons, $n$, in the considered diagram.

In the QGSM one calculates the invariant cross section
\begin{equation}
\frac{x_E}{\sigma_{inel}}\cdot\frac{d\sigma}{dx} = \int\frac{E}{\sigma_{inel}}\cdot\frac{d\sigma}{d^3p}\cdot d^2p_T \;,
\end{equation}
where $x=2p_{\|}/\sqrt{s}$ is the Feynman variable $x_F$, and
$x_E=2E/\sqrt{s}$, 
and one has then to use one value of $\langle p^2_T \rangle$ (here we have
taken the value $\langle p^2_T \rangle = 0.35$
(GeV/c)$^2$) to make the transition to the values of $d\sigma/dx_F$ which are presented in
the experimental papers~\cite{WA89}.

Thus, for the case of a nucleon target the inclusive 
spectrum of a secondary hadron $h$ in QGSM has the form~\cite{KTM}:
\begin{equation} 
\frac{x_E}{\sigma_{inel}}\cdot\frac{d\sigma}{dx}=\sum_{n=1}^{\infty}w_{n}\cdot
\phi^{h}(x,n) + w_D \cdot \phi_D^{h}(x)   \ .
\end{equation}
The functions $\phi^{h}(x,n)$ determine the contribution 
of diagrams with $n$ cut Pomerons, and $w_{n}$ is the probability for this 
process with $n$ cut Pomerons to occur~\cite{TM}. The second term in the right-hand 
side of Eq.~(3) describes the contribution of diffraction dissociation 
processes, where the triple-Reggeon diagrams are also included~\cite{KTM,2r}.  
The expressions of $\phi^{h}(x,n)$ for
$\Sigma N$ ($N=p,n$) collisions in Eq.~(3) have the form~\cite{KTM,Petya}:
\begin{equation}
\phi_{\Sigma N}^{h}(x,n) = f_{qq}^{h}(x_{+},n)\cdot f_{q}^{h}(x_{-},n) +
f_{q}^{h}(x_{+},n)\cdot f_{qq}^{h}(x_{-},n) +
2(n-1)f_{s}^{h}(x_{+},n)\cdot f_{s}^{h}(x_{-},n)\ \  ,
\end{equation}
\noindent where
\begin{equation}
x_{\pm} = \frac{1}{2}[\sqrt{4m_{T}^{2}/s+x^{2}}\pm{x}] \; ,
\end{equation}
with $m_T = \sqrt{m^2 + p^2_T}$ the transverse mass of the produced hadron,
and $f_{qq}$, $f_{q}$, and $f_{s}$ corresponding to the contributions
of diquarks, valence quarks, and sea quarks, respectively. The quantities
$f_i(x_+,n)$ and $f_i(x_-,n)$ account for the contributions to $\phi(x,n)$ of the
$\Sigma$-hyperon beam and of the target nucleon, and they are determined by the 
convolution of the diquark and quark distribution functions with the 
corresponding 
fragmentation functions, e.g.,
\begin{equation}
f_{qq}^{h}(x_{+},n) = \int_{x_{+}}^{1} u_{qq}(x_{1},n)\cdot
G_{qq}^{h}(x_{+}/x_{1}) \ dx_{1} .
\end{equation}

For the case of nuclear targets one has to consider the different possibilities
of one or several Pomeron cuts in each of the $\nu$ hadron-nucleon 
inelastic interaction blobs, as well as of cuts between Pomerons. For a 
$\Sigma^- A$ 
collision, one of the cut Pomerons links a valence diquark and a valence 
quark of the hyperon projectile with a valence quark and a diquark of one 
target nucleon, respectively, while the additional Pomerons link the sea quark-antiquark 
pairs of the projectile, either with diquarks and valence quarks, or with sea 
quark-antiquark pairs, of the target nucleons.

\begin{figure}[htb]
\centering
\vskip -1.5cm
\includegraphics[width=.5\hsize]{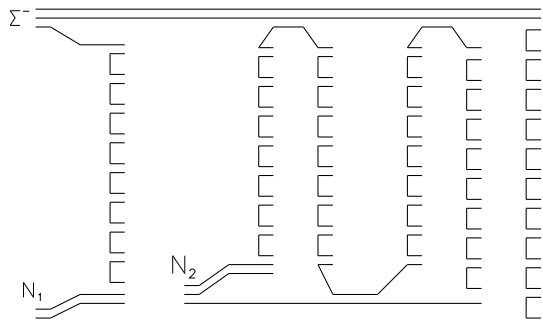}
\vskip -1.5cm
\caption{One of the diagrams for the inelastic interaction of an incident 
$\Sigma^-$-hype\-ron with two target nucleons $N_1$ and $N_2$ in a 
$\Sigma^{-} A$ collision.}
\label{af1}
\end{figure}

As one example, the diagram for the inelastic interaction of the $\Sigma^-$-beam with two
target nucleons is shown in Fig.~\ref{af1}. In the blob of the $\Sigma^- N_1$
inelastic interaction one Pomeron is cut, while in the blob of the 
$\Sigma^- N_2$ interaction two Pomerons are cut.
To include all diagrams, i.e.
to account for all possible Pomeron
configurations and permutations, is essential for a correct calculation.
The process shown in 
Fig.~\ref{af1} satisfies the condition that the absorptive part of the 
hadron-nucleus amplitude
is determined by combinations of the absorptive parts of hadron-nucleon
interactions, according to rules given in refs.~[16-19].

For $hA$ collisions, where $n$ inelastic interactions occur with $\nu$ target nucleons,
one has that $n \geq \nu$, and $1 \leq n_i \leq n-\nu+1$, $n_i$ being the number of cut Pomerons
connecting with the $i$-$th$ target nucleon. By denoting the relative weight of the
contribution with $n_i$ cut Pomerons in every $hN$ blob as $w^{hN}_{n_i}$, and by
using the same procedure as in ref.~\cite{KTMS}, we can write
the corresponding expressions for the inclusive spectrum of the secondary hadron $h$
produced in a $\Sigma A$ collision, where all possible Pomeron permutations and all possible
different quark contents of the protons and neutrons in the target
have to be accounted for.

%\begin{eqnarray}
%\frac{x_E}{\sigma^{prod}_{\Sigma A}}\cdot\frac{d \sigma}{dx_F} &=&
%\sum^A_{\nu=1} V^{(\nu)}_{\Sigma A}\cdot\left\{ \sum^{\infty}_{n=\nu}
%\sum^{n-\nu+1}_{n_1 = 1} \cdot \cdot \cdot
%\sum^{n-\nu+1}_{n_{\nu}=1} \prod^{\nu}_{l=1} w^{\Sigma N}_{n_l} \right. \nonumber\\
%&\cdot& [f^h_{qq}(x_+,n)\cdot f^h_q(x_-,n_l) +
%f^h_q(x_+,n)\cdot f^h_{qq}(x_-,n_l) \nonumber\\
%%& + & \sum^{2n-2}_{m=1} f^h_s(x_+,n)f^h_{qq,q,s}(x_-,n_m)]
%&+& \sum^{2n_l-2}_{m=1} f^h_s(x_+,n)\cdot f^h_{qq,q,s}(x_-,n_m)] \Bigr\} \;,
%\end{eqnarray}
In particular, the contribution to the inclusive spectrum of the diagram in 
Fig.~\ref{af1} is written as follows:

\begin{eqnarray}
\frac{x_E}{\Sigma^{prod}_{\Sigma A}}\cdot\frac{d \sigma}{dx_F} &=&
2 V^{(2)}_{\Sigma A}\cdot w^{\Sigma N_1}_1\cdot w^{\Sigma N_2}_2
\cdot\left\{ f^h_{qq}(x_+,3)\cdot f^h_q(x_-,1)\right. \nonumber \\
&+& f^h_q(x_+,3)\cdot f^h_{qq}(x_-,1) + f^h_s(x_+,3)\cdot[f^h_{qq}(x_-,2) +
f^h_q(x_-,2) \nonumber \\
&+& 2f^h_s(x_-,2)] \left. \right\} \;,
\end{eqnarray}
where $V^{(\nu)}_{\Sigma A}$ is the probability of pure inelastic
(non diffractive) interactions with $\nu$ target nucleons of a nucleus A to
occur.

The diquark and quark distributions, as well as the fragmentation functions,
are determined from Regge intercepts, and their expressions were presented in
Appendix 1 of ref. \cite{ACKS} (see also \cite{SJ1,AMS}). Now, for
the case of the presence of a $ds$ diquark in the beam, they are given in the 
Appendix of this paper
(see below).

For secondary baryon production, the diquark fragmentation function contains
two contributions. The first one corresponds to the production from the sea
of a $B\bar{B}$ pair in the midrapidity region (see Fig.~\ref{figabbc}), and it will
be discussed in detail in the Appendix.
\begin{figure}[htb]
\vskip -4.5cm
\hspace{4.cm}
\includegraphics[width=.8\hsize]{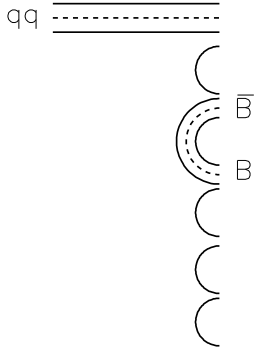}
\vskip -2.2cm
%%\vskip -0.25cm
\caption{Cut-chain diagram corresponding to the diquark fragmentation 
function for the production of a central $\bar{B}B$ pair.}
\label{figabbc}
\end{figure}

\begin{figure}[htb]
\centering
\vskip -3.8cm
\includegraphics[width=.55\hsize]{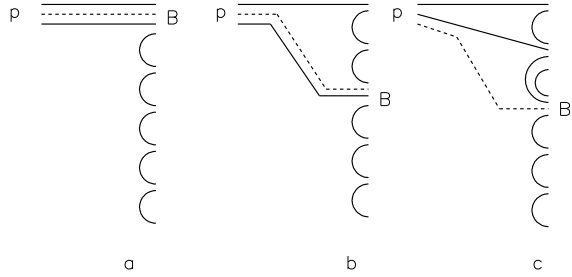}
\vskip -1.5cm
\caption{QGSM diagrams describing secondary baryon $B$ production by
diquark $d$: (a) initial SJ together with two valence quarks and one sea
quark, (b) SJ together with one valence quark and two sea quarks,
and (c) SJ together with three sea quarks.}
\label{f4}
\end{figure}

The second contribution is connected with the direct fragmentation of the
incident diquark into a secondary baryon with conservation of the SJ.
In the frame of QGSM three possibilities exist for
this second contribution~\cite{ACKS}. The secondary
baryon can consist of: (a) the SJ together with two valence and one sea quarks,
(b) the SJ together with one valence and two sea quarks, and (c) the SJ
together with three sea quarks. These three possibilities are shown in Fig.~\ref{f4}.
%\vspace*{-2.4cm}

The fraction of the energy of the incident baryon carried by the secondary
baryon decreases from (a) to (c), whereas the mean rapidity gap between
the incident and secondary baryon increases.

The processes shown in Figs.~\ref{f4}a and \ref{f4}b are the standard ones in
QGSM and DPM, and they determine the main contribution to the multiplicity of secondary
baryons in the fragmentation region.

On the other hand, the diagram shown in Fig.~\ref{f4}c
leads to the difference in baryon and antibaryon production at rapidities far from
the incident baryon (baryon charge diffusion in rapidity space). The role of
such a process in the description of experimental data was considered in
detail in refs. [21-26]. For $\Sigma A$ collisions at 345 GeV/c the relative
contribution of this diagram is rather small.

The fragmentation function of diquark $d$, with quark content $d=q_1q_2$,
into a secondary baryon $B$ through the processes shown in  Fig.~\ref{f4}a,
\ref{f4}b, and \ref{f4}c has the form~\cite{ACKS}:
\begin{equation}
G_{SJ}^{B}(z) = a_N\cdot z^{\beta}\cdot [ v_{q_1q_2}^B\cdot z^{2.5-\beta} + 
(v_{q_1}^B + v_{q_2}^B)\cdot z^{2 - \beta}\cdot (1-z) + v_0^B\cdot\varepsilon\cdot 
(1-z)^2]\cdot (1-z)^{\gamma} \;.
\end{equation}
Here $\beta=1-\alpha_{SJ}$, with $\alpha_{SJ}$ being the intercept of SJ Regge
trajectory, $\varepsilon$ is the relative suppression factor of the (c) contribution
with respect to the processes (a) and (b), and $a_B$ is a normalization
parameter. In the present calculations we use the values
$\varepsilon=0.024$, $\alpha_{SJ}=0.9$, and $a_N = 1.33$, as in
ref.~\cite{SJ1}. The factor $(1-z)^{\gamma}$ accounts for the fact that the
intercept of the $\phi$-meson Regge trajectory, $\alpha_{\phi}\approx 0$, is
smaller than the standard non-vacuum Reggeon intercept, $\alpha_R\approx 0.5$.
The value of the parameter $\gamma$ is half the difference
between the strangenesses of the considered diquark and that of the secondary baryon.
The powers of $z$ and $(1-z)$ are changed when either $q_1$ or $q_2$ is a strange quark.
The values of $v_i^B$ for different quarks i and baryons B are determined by quark
combinatorics~\cite{AS,CS}. These values are presented in the Appendix (see below).

\section{QGSM description of the experimental data}

The experimental data for $\Lambda$, $\Sigma^-$, $\Sigma^+$,
$\Xi^-$, and $\Omega^-$ baryons, $\bar{\Lambda}$ and $\overline{\Xi}^+$ antibaryons
production in $\Sigma^-$ collisions with C and Cu targets are
presented in ref.~\cite{WA89} in terms of $d\sigma/dx_F$.

\begin{figure}[htb]
\centering
%\vskip -1.cm
\includegraphics[width=.48\hsize]{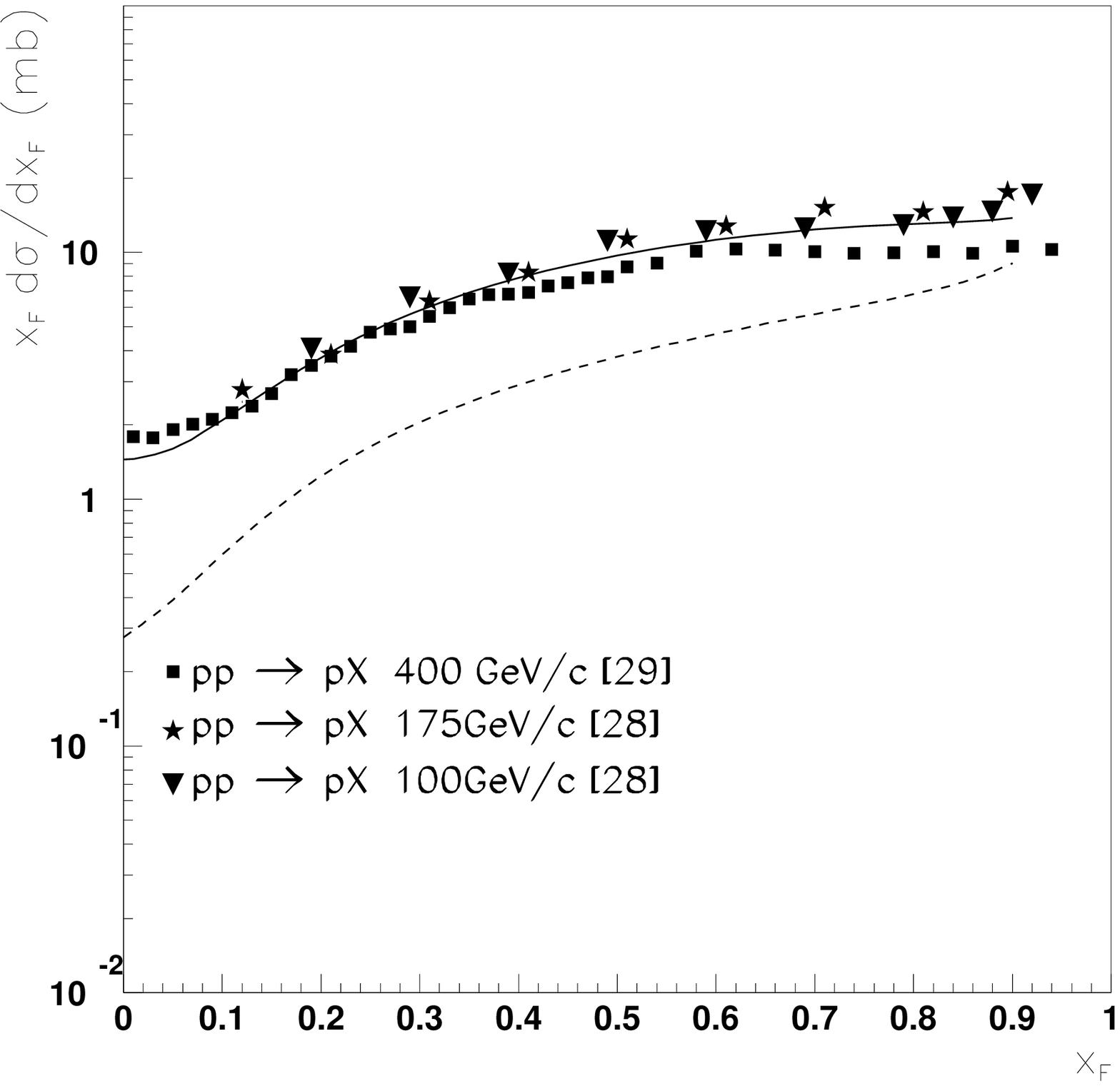}
\includegraphics[width=.48\hsize]{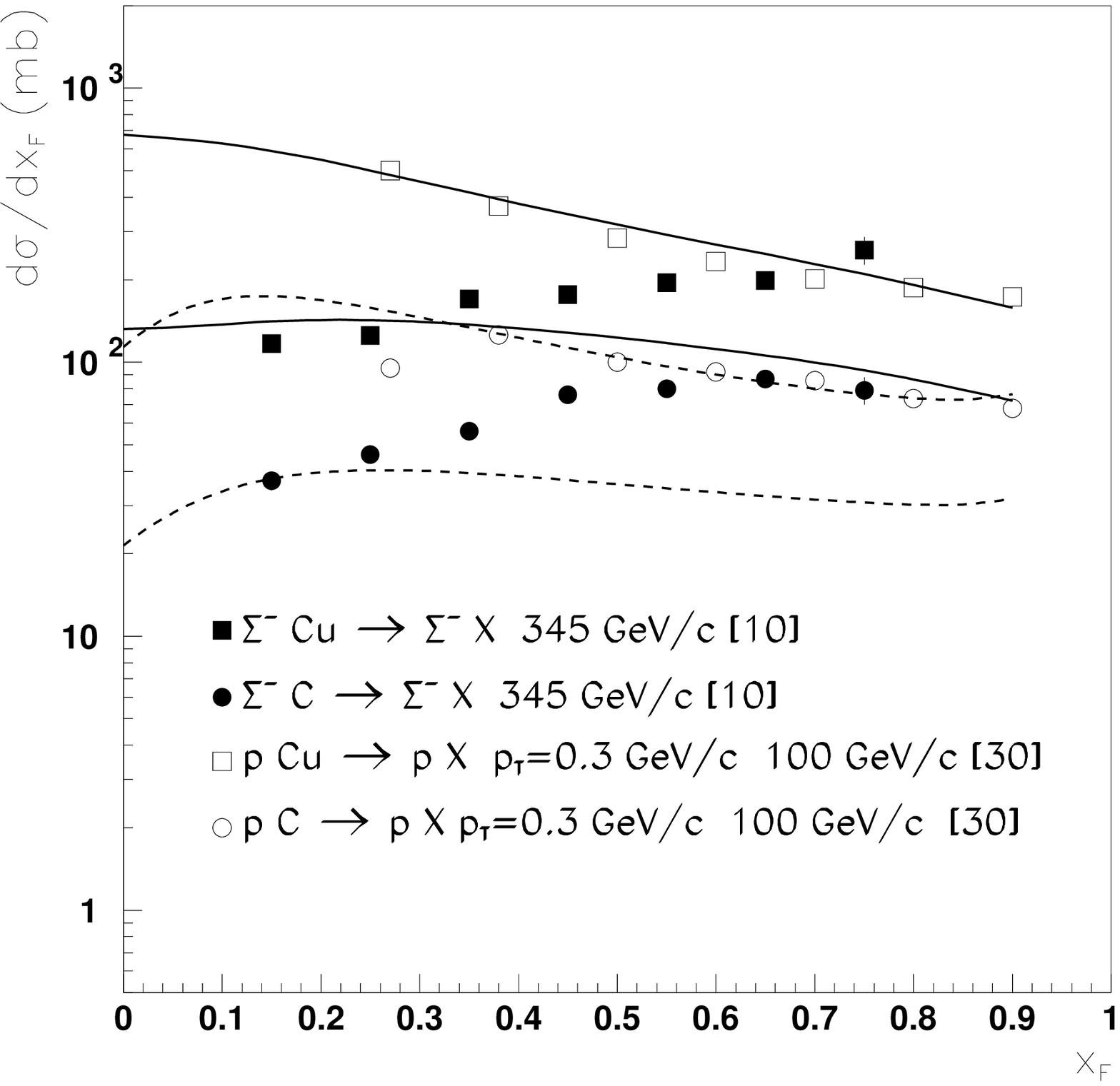}
\vskip -0.5cm
\caption{Experimental Feynman-$x$ distributions of secondary $p$ produced in $pp$
collisions at 100, 175, and 400 GeV/c, together with their description by QGSM,
and the QGSM predictions for $\Sigma^-p \to \Sigma^- X$ (dashed curve), left panel.
Experimental Feynman-$x$ distributions of secondary $\Sigma^-$
produced in $\Sigma^-C$ and $\Sigma^-Cu$ interactions at 345 GeV/c \cite{WA89},
and similar distributions of secondary $p$ on the same targets~\cite{barton},
and their comparison with the the predictions for $\Sigma^-$'s
(solid curves) and for protons (dashed curves) by QGSM, right panel.
}
\label{sigmaminpp}
\end{figure}

Let us start our analysis from the left panel of Fig.~\ref{sigmaminpp}, where the data
on $p$ spectra at $pp$ collisions and their comparison with our old QGSM calculations are shown.
The agreement is good, as it was usually obtained in our previous papers.

For comparison, we also present similar predictions for the case
$\Sigma^-p \to \Sigma^- X$, which is theoretically similar to $pp \to pX$, but about 2 times smaller
in the region of moderate $x_F$. It's generally accepted that this
difference can be connected to the rather large probability to
produce a $\Lambda$ in the $\Sigma^-$ case. At small $x_F$ one also has suppression of the $\Sigma^-$
production by sea quarks, what leads to a larger difference in the two considered reactions.

The experimental data on secondary protons production on nuclear targets
at 100 GeV/c and $p_T=0.3$GeV/c \cite{barton} are also in good agreement with 
the QGSM, as it is shown in right panel of Fig.~4 and in \cite{KTMS}. 
On the contrary, experimental data for $\Sigma^-$ production on $C$ and $Cu$ targets shown in
the right panel of Fig.~\ref{sigmaminpp} are in contradiction with our calculations when we use
the fragmentation functions directly taken from the Reggeon counting rules, as it was
done in all previous papers \cite{KTM,KaPi,KTMS,Sh}. 
One can immediately see that these data have different shape 
that the proton data for the same targets.
Unfortunately, 
the proton data \cite{barton} presented here were measured at fixed 
$p_T=0.3$GeV/c, and there are no proton production data on nuclear targets 
integrated over $p_T$, which could be used for direct comparison. However, the 
difference in the shapes of the distributions seems to be too large, especially 
when keeping in mind that the data of \cite{barton} were successfully described 
by the QGSM in ref. \cite{KTMS}. 

Now, let us compare the spectra of secondary $\Sigma^-$ and $\Lambda$ obtained 
in ref.~\cite{WA89}. The calculated ratio of these spectra, together with the
corresponding experimental points obtained by using the data of ref.~\cite{WA89},
are presented in Fig.~5. Both  $\Sigma^-$ and $\Lambda$ should be produced
in the interval $x_F=0.3$-$0.8$ mainly by the process shown in Fig.~\ref{f4}a, 
when the $ds$ diquark fragments into them by picking up, either an $u$ or a 
$d$-quark from the sea, respectively. 
The fragmentation functions for both channels should be somehow similar, with
differences coming from other contributions including triple-Reggeon
terms. This comes from the fact that $ds$ and 
$dd$ diquarks appear in the incident $\Sigma^-$ with probabilities 2/3 and 
1/3, respectively, and, on top of that, the fragmentation of the $dd$ diquark
into strange 
secondary baryons is suppressed by the strangeness suppression factor $S/L$ 
(see Appendix). So, the ratio $\Sigma^-$/$\Lambda$ should be approximately
constant in the considered interval. However, the experimental ratio of 
$\Sigma^-$/$\Lambda$ yields increases more than 3 times in the interval 
$x_F=0.3$-$0.6$, in total disagreement with our theoretical expectation. 
We want to stress that we have never met
such a large disagreement with the experimental data in any of our previous calculations,
and one has also to note that this disagreement with the experimental data on the ratio
$\Sigma^-$/$\Lambda$ can not be corrected by any theoretically meaningful modification
of the $ds$-diquark fragmentation functions.
\begin{figure}[htb]
\centering
%\vskip -1.cm
\includegraphics[width=.48\hsize]{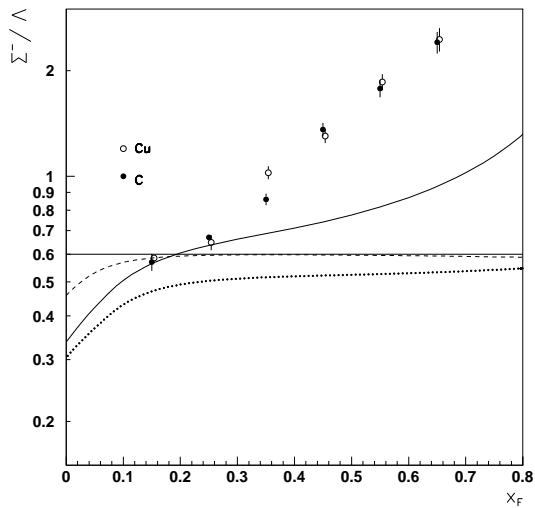}
\vskip -0.5cm
\caption {Experimental ratios of secondary $\Sigma^-$ to $\Lambda$ yields vs $x_F$
in $\Sigma^-C$ and $\Sigma^-Cu$ interactions at 345 GeV/c $\cite{WA89}$, 
together with the QGSM predictions for these ratio (solid curve). The 
dashed and dotted curves show the ratios when taking into account different contributions to
the QGSM calculations (see details in the main text).
}
\label{rat}
\end{figure}

The absolute value of $\Sigma^-$ to $\Lambda$ ratio in the QGSM is more
model dependent. In the simplest approximation, when all final $dds$ states 
are assumed to be $\Sigma^-$ and all $uds$ states are assumed to be $\Lambda$ 
(secondary $\Sigma^0$ are usually registered as $\Lambda$ after 
radiative $\Sigma^0 \to \Lambda \gamma$ decay), the ratio 
$\Sigma^-$/$\Lambda$ is equal to $1$. However, this can be changed by the 
resonance production, e.g. 
the state $\Sigma^*(1385)$, that can be produced with rather 
large cross section, has a dominant decay mode $\Lambda \pi$ that 
would transfer some part of $dds$ states into additional $\Lambda$'s, and
it would thus 
decrease the $\Sigma^-$/$\Lambda$ ratio.

To account for this effect we assume that when only $ds$ diquark
fragmentation is considered, the hyperon
production leads to the following empirical rule for the
$\Sigma^-$/$\Lambda$ ratio~\cite{CS}:
fragmentation is considered, the $\Sigma^-$/$\Lambda$ ratio would be.
\begin{equation}
R^{(ds)}_{\Sigma^-/\Lambda} = \Sigma^-/\Lambda = 0.6 \;.
\end{equation}
Such an assumption has provided a reasonable description of the
$\Lambda$ $x_F$ spectra in $pp$, $p\pi$, $kp$ collisions~\cite{ACKS,SJ1,AMS}.
Both $\Sigma^-$ and $\Lambda$ spectra should be affected by nuclear effects 
in a similar way.

To show the structure of our calculations in more detail, in Fig.~5 we present 
the ratio of $\Sigma^-$ to $\Lambda$ when only part of all 
contributions is accounted for. The dashed curve shows the result of calculations 
when only the first term in the right-hand side of Eq.~(4) is considered. In this 
case the ratio of $\Sigma^-$ to $\Lambda$ in the fragmentation region 
($x_F \geq 0.2$) is very close to 0.6, the very small difference from this value 
coming from the contribution of $dd$ diquark fragmentation. At small $x_F$, 
the values of $x_-$ in Eq.~(5) increase, leading to the decrease of the 
functions $f^n_q(x_-,n)$ in Eq.~(4), and to the decrease of the $\Sigma^-/\Lambda$ 
ratio.  

The ratio of $\Sigma^-$ to $\Lambda$ production when only fragmentation
of $ds$ diquark is considered, and without diffraction dissociation nor 
triple-Reggeon contributions, is shown in Fig.~5 by a dashed curve. If
all diquark and quark terms in Eq.~(4) are included (dotted curve), the
ratio is smaller and the agreement with the experimental data becomes worse,
indicating that the $ds$ diquark is responsible for the value $0.6$ of the
$\Sigma^-$ to $\Lambda$ production ratio.
The solid curve shows the result of the complete 
QGSM calculation with diffraction dissociation and triple-Reggeon 
contributions, and though now the theoretical curve goes up at large $x_F$,
it still presents a very significant disagreement with the experimental
data. It seems clear that the agreement of the QGSM 
predictions with the experimental data can not be obtained by simply
considering a slight 
variation of any of the contributions in Eq.~(4).

As the QGSM can not reproduce the experimental ratio of $\Sigma^-$ to $\Lambda$ 
production, it can not describe separately
the $x_F$-distributions of both
$\Sigma^-$ and $\Lambda$, that present clearly different experimental behaviours.
It is nevertheless interesting to find out
which one, $\Sigma^-$ or $\Lambda$ distributions, or both, are at the origine of the
disagreement in the ratio. The two $x_F$ distributions for $\Sigma^-$ and $\Lambda$
are presented in Fig.~6. 
\begin{figure}[htb]
\centering
%\vskip -1.cm
\includegraphics[width=.48\hsize]{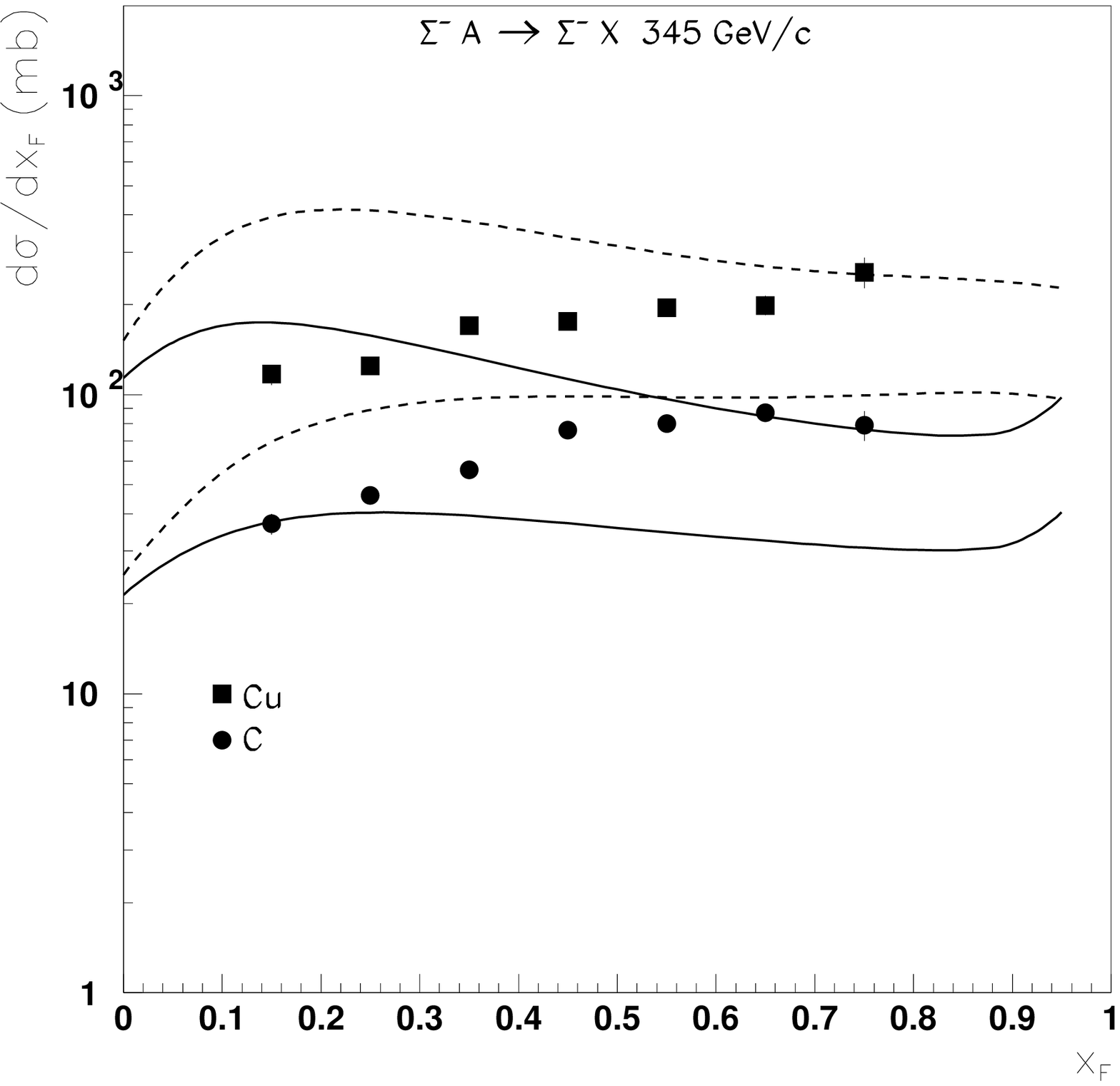}
\includegraphics[width=.48\hsize]{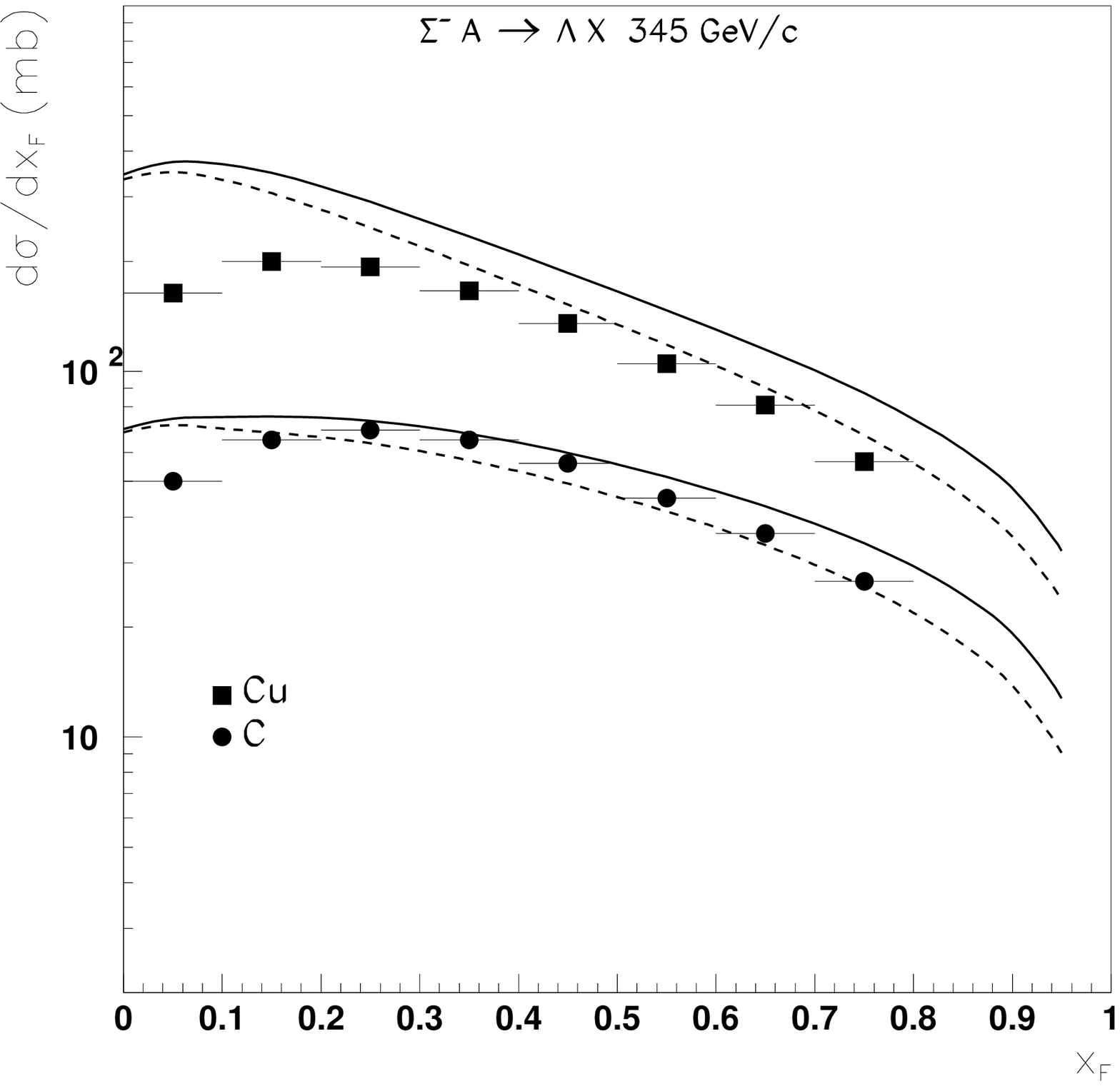}
\vskip -0.5cm
\caption {Experimental $x_F$ distributions of secondary $\Sigma^-$'s (left 
panel) and $\Lambda$'s (right panel), produced in $\Sigma^-C$ and $\Sigma^-Cu$ 
interactions at 345 GeV/c $\cite{WA89}$, together with the corresponding QGSM predictions. 
The solid curves show the result of the calculations with only the standard polynomial
terms. The dashed curves
in the left panel show the calculations with one additional polynomial
factor in the third term of the $ds$
fragmentation function, and the dashed curves in the right panel show the 
result of the calculations when accounting for the resonance contribution to the
spectra of $\Lambda$.}
\label{sigmaminpl}
\end{figure}

The difference between the $\Sigma^-$ and the $\Lambda$ $x_F$ distributions could be explained
by assuming that in the case of $\Lambda$
production, but not in that of $\Sigma^-$ production, the resonance decay
contribution is quite significant. In all cases, when including the resonance decay
contribution for $\Sigma^-$ production the $x_F$ distribution would become softer,
i.e. narrower, and consequently
the agreement with the experimental data would be worse.

The quark and diquark distribution and fragmentation functions are given 
by the Reggeon counting rules. Thus, for some fragmentation function 
$G^h_{qq}(z)$ in Eq.~(6) having asymptotical behaviours  $G^h_{qq}(z \to 0) \sim z^{\alpha}$ 
and $G^h_{qq}(z \to 1) \sim (1-z)^{\beta}$, the fragmentation function is
written as the simplest interpolation of these two asymptotical behaviours in the form~\cite{Kai}:
\begin{equation}
G^h_{qq}(z) = a_h\cdot z^{\alpha}\cdot (1-z)^{\beta} \;,
\end{equation}
where $a_h$ is a parameter which determines the inclusive density of a
produced hadron $h$. We will call such a form of the fragmentation function as the standard one.
However, a slightly more complicate form with additional polynomial factors and new 
parameters $b$ and $b_n$ is also possible:
\begin{equation}
G^h_{qq}(z) = a_h\cdot z^{\alpha}\cdot (1-z)^{\beta}\cdot (1 + b\cdot z^{b_n}) \;.
\end{equation}

The theoretical $x_F$ distributions of $\Sigma^-$ and $\Lambda$ shown by solid curves in Fig.~6
have been calculated by using
the standard form of the QGSM fragmentation funactions, without any additional polynomial factor
in Eq.~(11). The distributions so obtained 
are in an evident disagreement with the data for the case $\Sigma^-$ production, and the 
attempt to include an additional polynomial factor in the fragmentation function of 
$ds$ diquark (dashed curves in the left panel of Fig.~6) does not lead to any significant 
improvement of the agreement with the data, since though the absolute values of the spectra increase 
the shapes remain being wrong. 

In the case of $\Lambda$ production (right panels of Fig.~6), the agreement with the experimental data
of the QGSM calculation with the standard form of the fragmentation functions with the experimental
data is reasonable, and it becomes better
when taking into account that some part of the $\Lambda$-hyperons are produced after resonance 
decay and they consequently have smaller $x_F$. To include this effect in our calculation we 
have introduced into the fragmentation function Eq.~(\ref{dsL})) the additional 
factor $(1-\frac {z}{3})$. The QGSM results obtained with this modified fragmenation functions
are shown in Fig.~6 by dashed curves.   

The spectra of secondary $\Sigma^+$ and $\Xi^-$ are shown on Fig.~7. These 
secondaries are produced with cross sections several times smaller than 
secondary $\Sigma^-$ and $\Lambda$.

\begin{figure}[htb]
\centering
%\vskip -2.cm
\includegraphics[width=.48\hsize]{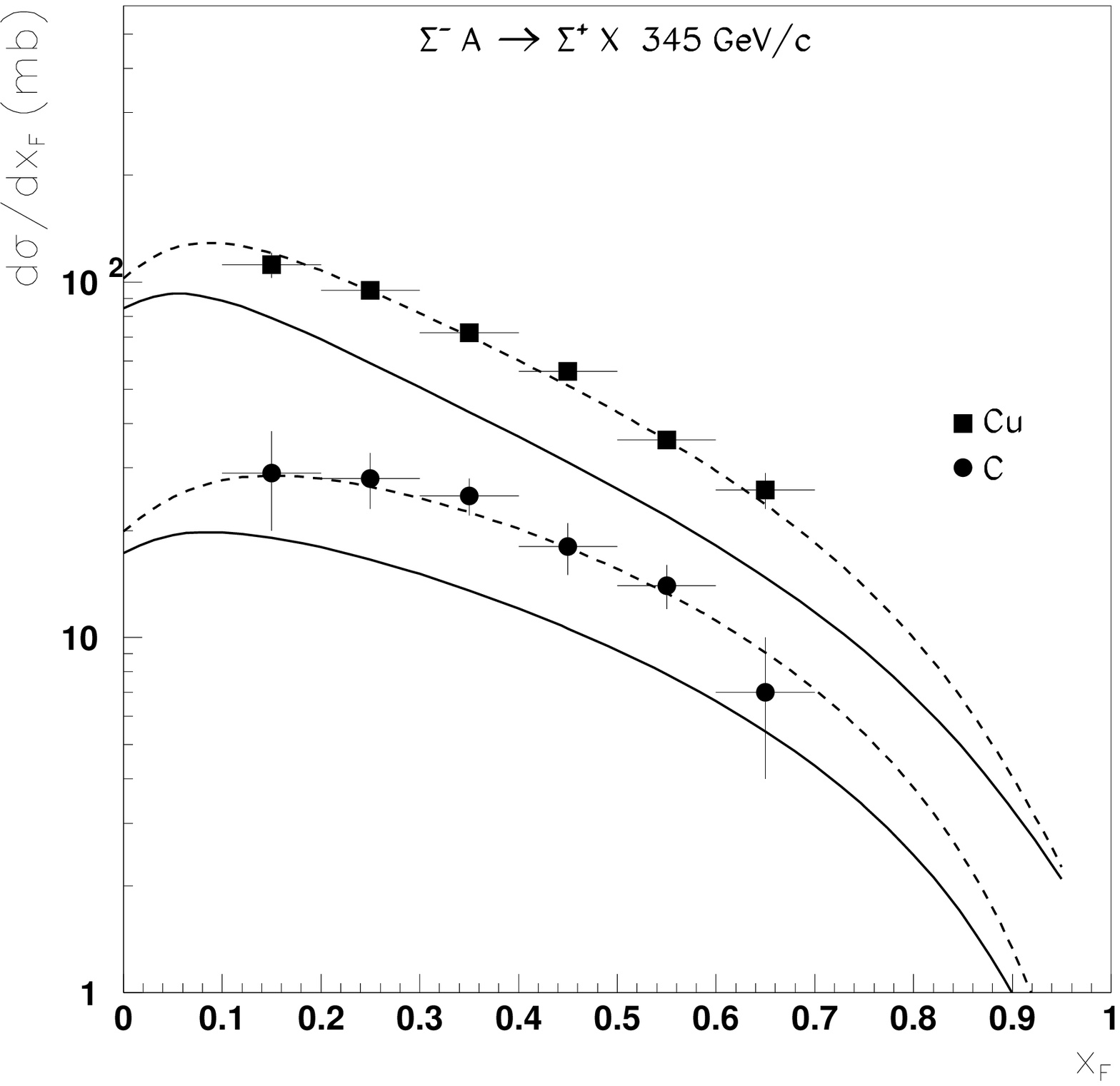}
\includegraphics[width=.48\hsize]{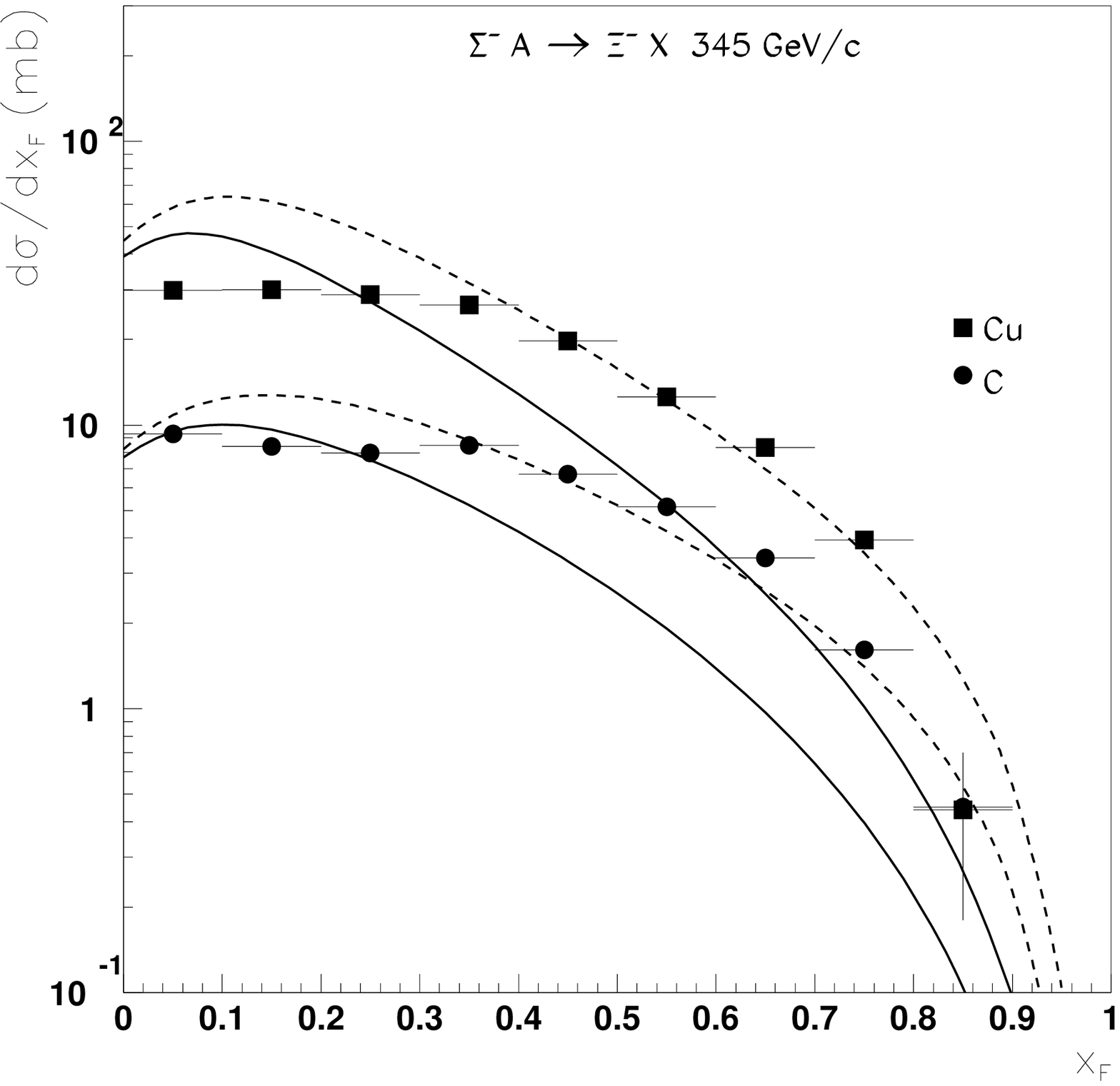}
\vskip -0.5cm
\caption{
Experimental $x_F$ distributions of secondary $\Sigma^+$ (left panel)
and $\Xi^-$ (right panel), produced in $\Sigma^-C$ and in $\Sigma^-Cu$
interactions at 345 GeV/c~\cite{WA89}, together with the corresponding QGSM
predictions. The solid curves show the result of the calculations with only
the standard QGSM expressions. The dashed curves are obtained with
with one additional polynomial
factor in the fragmentation functions.}
\label{n1ccul}
\end{figure}

In the standard approximation of the QGSM,
for $\Sigma^+$ production
only one valence $s$ quark, Eq.~(A.14), as well as an $s$ quark through
the diagram in Fig~3b, from the incident $\Sigma^-$ can be used to
fragment into $\Sigma^+$, but another possibility that is usually
considered is the resonance production of $\Sigma^{*0}(1385)$
or $\Lambda(1405)$ in the process of $ds$-diquark fragmentation
(see Fig.~3a), and the subsequent decay into $\Sigma^+ \pi^-$.
The results of the calculations in this approximation (shown by solid curves
in the left panel of Fig.~7) are in reasonable agreement with the
experimental data. The shapes
of the curves are correct, though the normalization are underestimated on the
level of 30\%. This disagreement can be corrected (dashed curves in the left
panel of Fig.~7) by introducing a polinomial factor $(1 + 5\cdot z)$ into the
last term of the $ds$ diquark fragmentation function in Eq.~(\ref{dssigpl}).

For the case of $\Xi^-$ production (right panel of Fig.~7) the contribution 
of the $ds$-diquark fragmentation is decreased by the strangeness production 
suppression factor. The calculation of $\Xi^-$ production with only the 
standard terms in the diquark fragmentation functions (solid curves in the
right panel of Fig.~7) results in a too fast decrease of the spectra
when increasing $x_F$. The calculation with an additional polinomial
factor $(1 + 3\sqrt{z})$ in the last term of the diquark fragmentation
function Eq.~(A.38) leads to a better agreement with the experimental
data (dashed curves on the right panel of Fig.~7), except for the region of
low $x_F$, where the model results are significantly higher than the experimental
data.

In the case of secondary $\Omega^-$ production by $\Sigma^-$ beam, the incident $s$ quark should pick up two
strange quarks from the sea. The cross section of this process should clearly be small due to the presence of
the squared strangeness suppression factor. The experimental points for secondary $\Omega^-$ production are
presented in Fig.~8. Here the standard QGSM predictions are in reasonable agreement with the data.
\begin{figure}[htb]
\centering
%\vskip -2.cm
\includegraphics[width=.4\hsize]{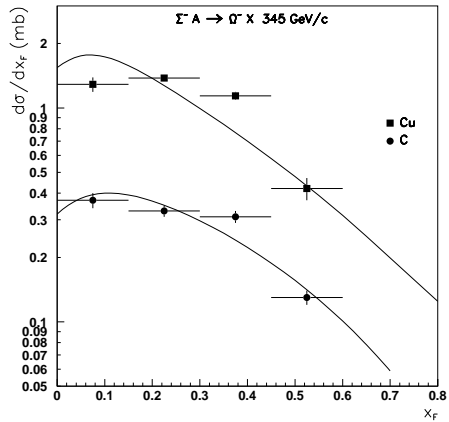}
\vskip -0.5cm
\caption{The $x_F$ distributions of secondary $\Omega^-$'s produced in $\Sigma^-C$
and $\Sigma^-Cu$ interactions at 345 GeV/c \cite{WA89}. The curves show the QGSM predictions
with only the standard QGSM expressions in the diquark fragmentation functions.}
\label{n1ccull}
\end{figure}

\begin{figure}[htb]
\centering
\vskip -0.5cm
\includegraphics[width=.49\hsize]{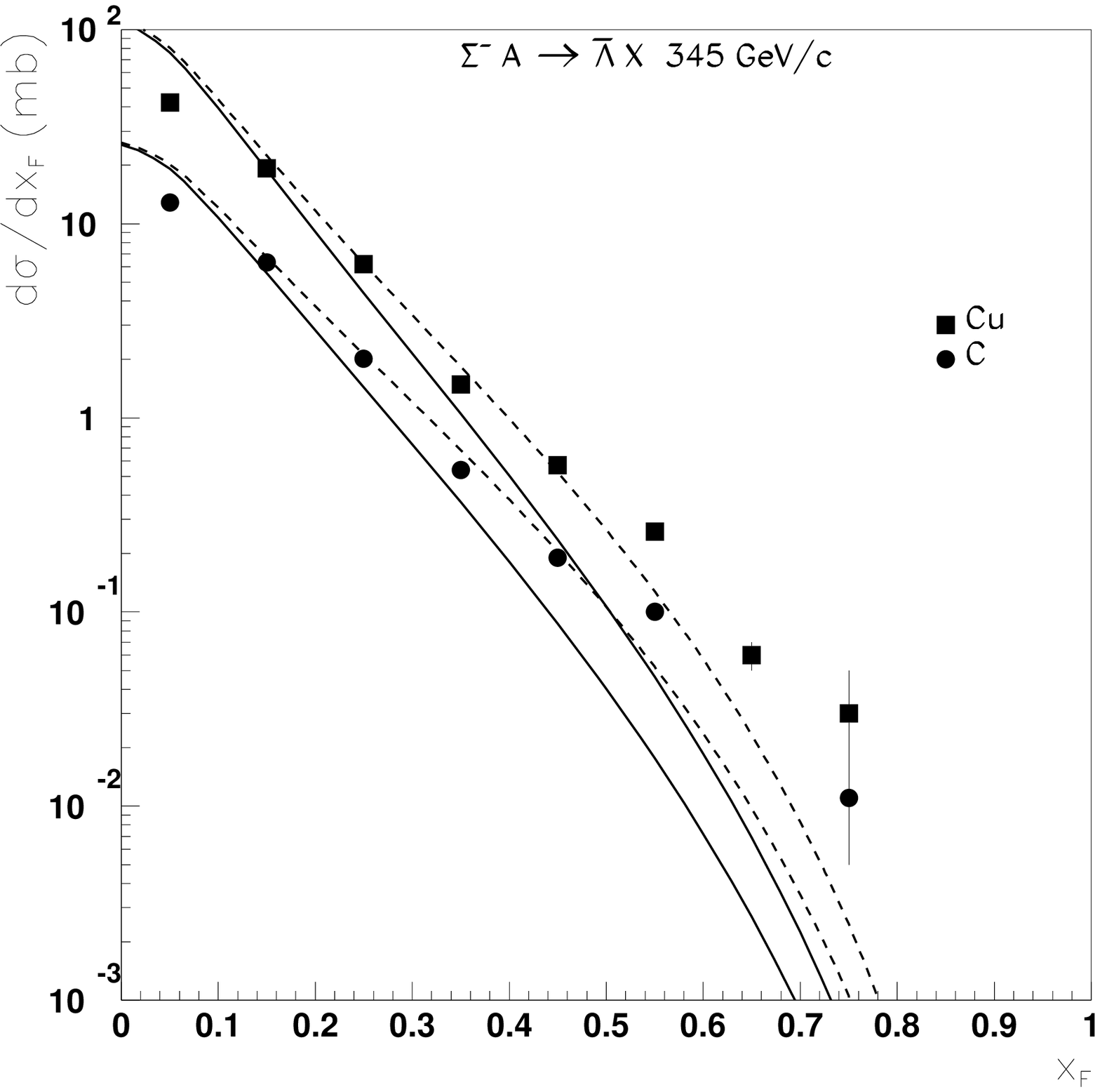}
%%%\vskip -1.cm
\includegraphics[width=.49\hsize]{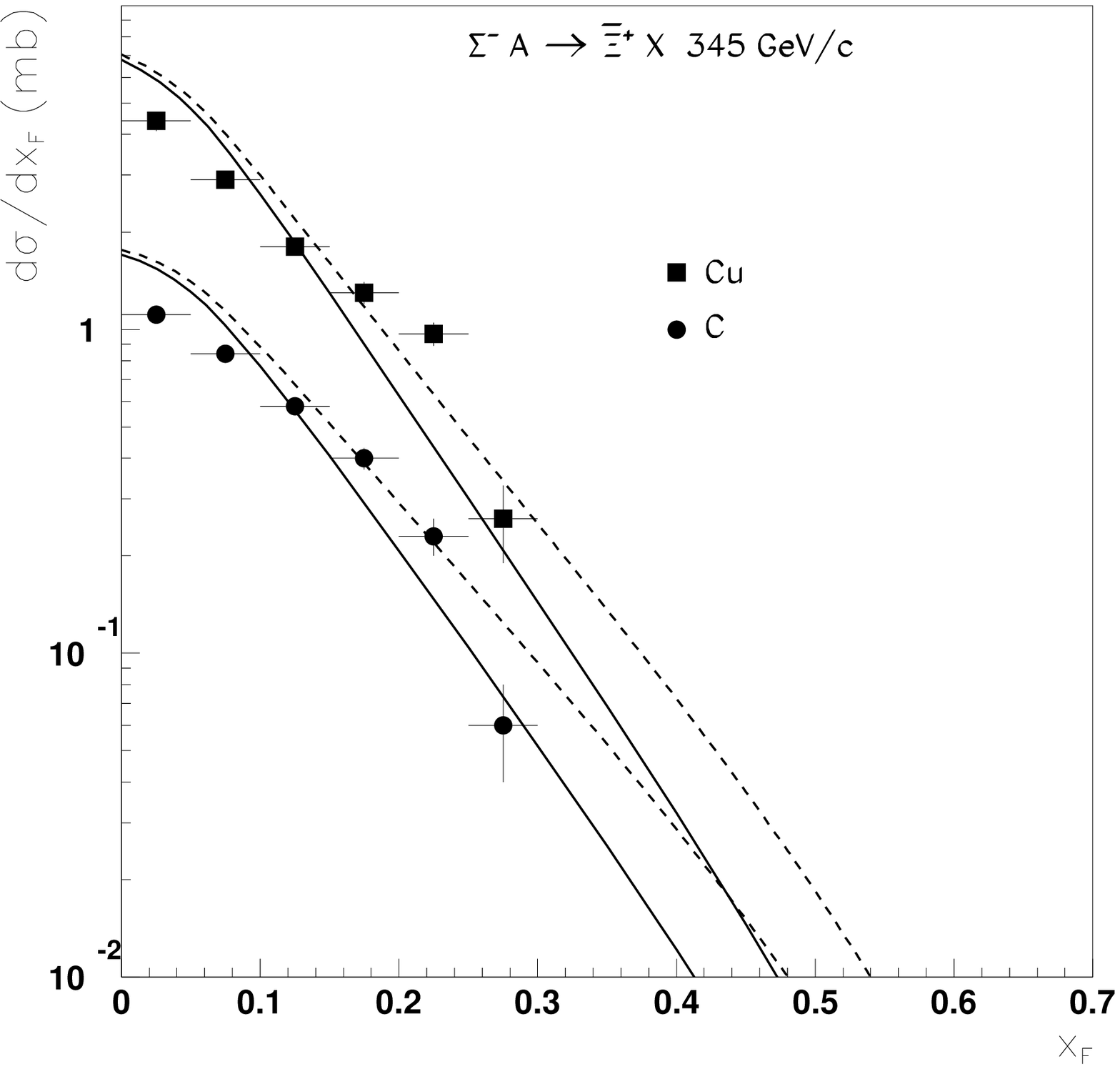}
\vskip -0.5cm
\caption{The $x_F$ distributions of secondary $\bar{\Lambda}$ (left panel) 
and $\bar{\Xi}^-$ (right panel) produced in $\Sigma^-C$ and in $\Sigma^-Cu$ 
interactions at 345 GeV/c, together with the corresponding standard QGSM 
calculations (solid curves) and with the QGSM calculations obtained by
including an additional polynomial factor into 
the $ds$ fragmentation functions of eqs.~(A.20) and (A.21).}
\label{n1ccual}
\end{figure}

The experimental yields of $\bar{\Lambda}$'s and $\overline{\Xi}^+$'s~\cite{WA89}, which
only contain sea antiquarks and do not depend on the SJ contribution, are shown
in Fig.~\ref{n1ccual}. The corresponding description by the standard QGSM fragmentation
functions clearly underestimates the $\bar{\Lambda}$ yields at $x_F > 0.4$.
For the case
of $\overline{\Xi}^+$ production the agreement is reasonable.
The description of the
$\bar{\Lambda}$ and $\overline{\Xi}^+$ spectra obtained by using the fragmentation functions 
of eqs.~(A.20) and (A.21), where an additional polinomial factor $(1 + 20\cdot z^2)$ has been
included, leads
to a better agreement with the experimental data of the $\bar{\Lambda}$ spectra, letting
apart some overestimation in the small $x_F$ region.
\vskip -0.5cm

\vskip -0.75cm

\section{Conclusions}
\vskip -0.2 truecm

We present the QGSM description of the experimental data~\cite{WA89} on 
secondary hyperon production in $\Sigma^-C$ and $\Sigma^-Cu$ collisions at 
345 GeV/c. These data are of special interest because the main contribution 
to the spectra of secondary $\Lambda$, $\Sigma^-$, and $\Xi^-$ at 
$x_F \geq 0.3$ comes from the direct fragmentation Fig.~3a of the 
incident $ds$ diquark, with rather small background from another subprocesses. 

From the invariant cross section in Eq.~(3)
and by using the value $\langle p^2_T \rangle = 0.35$
(GeV/c)$^2$, we get the values of $d\sigma/dx_F$ that could be compared to those
presented in the experimental papers~\cite{WA89}.
In all cases, except for $\Sigma^+$ production, we overestimate the experimental
data at small $x_F$.

In the region $x_F > 0.3$-$0.4$, practically all the experimental $x_F$ distributions
measured by the WA89
Collaboration~\cite{WA89} are wider than the corresponding QGSM predictions
obtained by using the standard diquark fragmentation functions given by Reggeon
counting rules. To solve this steady disagreement, we have included in the diquark 
fragmentation
functions (mainly in $ds$ diquark fragmentation) one additional polynomial factor
which takes the spectra up at $x_F > 0.3$-$0.4$.
Another problem faced when trying to
describe the experimental data of ref.~\cite{WA89} is connected with the behaviour
at small $x_F$. The peaks present in the theoretical estimations and shown in
Fig.~\ref{n1ccull} are the natural result of dividing the rather flat function
$x_E\cdot\frac{d\sigma}{dx}$ by $x_E$. The experimental behaviours of $d\sigma /dx$
correspond, on the contrary, to deep minima in $x_E\cdot\frac{d\sigma}{dx}$.

Nevertheless, the most disturbing question we face when comparing the QGSM
predictions for the $x_F$ spectra of secondary hyperons comes from the fact that
the experimental $x_F$ distributions of secondary 
$\Sigma^-$ are very different to the corresponding $x_F$ spectra 
of secondary protons in $pA$ collisions. Since the standard QGSM predictions have
always provided a good agreement with the experimental data on the $x_F$ spectra
of secondary non-strange secondaries, from the theoretical
point of view it would be puzzling if the disagreement of the corresponding
spectra of strange secondaries with the experimental data would be confirmed.
On top of that, should the theoretical predictions
for the $\Sigma^-$ case be modified to agree with the currently available experimental data,
this would mean significant changes in the standard QGSM diquark distributions
and/or in the $ds$-diquark fragmentation function to $\Sigma^-$. However, one actually has not
any apparent theoretical support for those changes, and that could even result in the violation of the
SU$(3)$ flavour symmetry. Let's then see whether the experimental data on the $x_F$ spectra
for secondary $\Sigma^-$ will be confirmed or corrected in the future.
\vskip -0.1 truecm

\newpage

%\vskip 0.2cm
{\bf Acknowledgements}
 
This paper was supported by Ministerio Educaci\'on y Ciencia of Spain under 
the Spanish Consolider-Ingenio 2010 Programme CPAN (CSD2007-00042), by Xunta de Galicia
(Galiza, Spain)
under project FPA2008-01177, and also by Universidade de Santiago de Compostela
through grants RFBR-07-02-00023 and RSGSS-1124.2003.2.

%\newpage

\setcounter{equation}{0}

\vspace{-0.35cm}
\section*{Appendix}
\vspace{-0.1cm}

The diquark and quark distribution functions in $\Sigma^-$-baryon for a
diagram with $n$ cut Pomerons have been parametrized as follows:
\begin{eqnarray}
u_{dd}(x,n) &=& C_{dd}\cdot x^{\alpha_{R} - 2 \alpha_{B}}\cdot (1 - x)^{-\alpha_{\varphi} +m_2}\;, \\
u_{ds}(x,n) &=& C_{ds}\cdot x^{\alpha_{R} - 2 \alpha_{B}+(\alpha_{R}(0)-\alpha_{\varphi}(0))}
\cdot (1 - x)^{-\alpha_{R} + m_1}\;, \\
u_{d}(x,n) &=& C_{d}\cdot x^{-\alpha_{R}}\cdot (1 - x)^{\alpha_{R} - 2 \alpha_{B} + 
(\alpha_{R}(0)-\alpha_{\varphi}(0))+ m_1}\;, \\
u_{s}(x,n) &=& C_{s}\cdot x^{-\alpha_{\varphi}}\cdot (1 - x)^{\alpha_{R} - 2 \alpha_{B}+ m_2}\;, \\
u_{sea}(x,n) &=& C_{s}\cdot x^{-\alpha_{R}}\cdot (1 - x)^{\alpha_{R} - 2 \alpha_{B} + n - 1+k}\;.
\end{eqnarray}
The values of $m_1$  and $ m_2$ in Eqs.~(A.1)-(A.4) are determined from 
momentum conservation:
%\bc
\begin{eqnarray}
\langle x\rangle_d + \langle x\rangle_{ds} + 2\cdot(n-1)\cdot\langle x\rangle_{sea} &=& 1
\;, \\
\langle x\rangle_s + \langle x\rangle_{dd} + 2\cdot(n-1)\cdot\langle x\rangle_{sea} &=& 1
\;,
\end{eqnarray}
where   
\be
\langle x\rangle_i =\int_{0}^{1} x\cdot u_i(x)\cdot dx \; , 
\ee
and
\be
\int_{0}^{1} u_{i}(x,n)\cdot dx = 1,\hspace{0.25cm} i=q,qq,sea \;.
\ee
For the values $\alpha_R = 0.5$, $\alpha_{\phi} = 0$, $\alpha_B = -0.5$, and $k = 0$,
we obtain $m_1 = m_2 = \frac76(n-1)$.  

The fragmentation functions of quarks and diquarks used for 
the description of strange baryon inclusive spectra are presented here. 

For quarks one has:

\begin{eqnarray}
G_{s}^{\Lambda}(z) &=& a_{\bar{N}}\cdot (1 - z)^{\lambda +
\alpha_{R}-2\cdot \alpha_{B}}\cdot (1 + a_{1}z^{a_{1n}}) \;, \\
G_{s}^{\bar{\Lambda}}(z) &=& a_{\bar{\Lambda}}\cdot (1 - z)^{\lambda+ \alpha_{R}
-2\alpha_{B} + 2(1 - \alpha_{R}) + 2\Delta \alpha} \cdot (1 + a_1 z^{a_{1n}}) \;, \\
%\label{u4spl}
G_{u}^{\Sigma^+}(z) &=& G_d^{\Sigma^-}(z) = a_{\bar{\Sigma}^{\pm}}\cdot 
(1-z)^{\lambda + \alpha_R - 2 \alpha_B+\Delta \alpha }\cdot (1+a_{1}z^{a_{1n}}) \;, \\
%\label{u4sm}
G_{u}^{\Sigma^-}(z) &=& G_d^{\Sigma^+}(z) = a_{\bar{\Sigma}^{\pm}}\cdot
(1-z)^{\lambda + \alpha_R - 2 \alpha_B + 2(1- \alpha_R) + \Delta \alpha}
\cdot (1+a_{1}z^{a_{1n}}) \;, \\
G_{s}^{\Sigma^-}(z) &=& G_{s}^{\Sigma^+}(z) = \frac{a_{\bar{\Sigma}}}
{a_{\bar{\Lambda}}}\cdot G_{s}^{\Lambda}(z) \;, \\
%\label{tksi}
G_{s}^{\Xi^-}(z) &=& a_{\overline{\Lambda}}\cdot (1 - z)^{\lambda+ \alpha_{R}
-2\alpha_{B} + \Delta \alpha}\cdot (1 + a_{1}z^{a_{1n}})\;, \\
%\label{taksi}
G_{s}^{\bar \Xi^+}(z) &=& a_{\overline{\Xi}}\cdot (1 - z)^{\lambda+ \alpha_{R}
-2\alpha_{B} + 3\Delta \alpha+2(1-\alpha_R)}\cdot (1 + a_{1}z^{a_{1n}})\;, \\
%\begin{equation}
%\label{taomega}
%G_{s}^{\bar\Omega}(z) = a_{\overline{\Omega}}(1 - z)^{4\lambda+ \alpha_{R}
%-2\alpha_{B} + 1.}\cdot (1 + a_{1}z^2) \;, 
%\end{equation}
%\label{taso}
G_{s}^{\Omega}(z) &=& a_{\overline{\Xi}}\cdot(1 - z)^{\lambda+ \alpha_{R}
-2\alpha_{B} + 2\Delta \alpha}\cdot (1 + a_{1}z^{a_{1n}}) \;,
\end{eqnarray}
where $\Delta \alpha = \alpha_{R} - \alpha_{\phi}$, $a_1$ = 12, $a_{1n}$ = 2, and
$\lambda=2\alpha^{\prime}\cdot <p_{t}^2>=0.5$. The fragmentation functions 
which are not presented here can be found in ref. \cite{ACKS}. The values of 
the parameters $a_{\bar{\Lambda}}$, $a_{\bar{\Sigma }}$, $a_{\bar{\Xi}}$, and 
$a_{\bar{\Omega}}$ can be obtained from $a_{\bar{N}}$ through quark combinatorics
(see below).

The QGSM diquark fragmentation functions corresponding to
antibaryons production through the diagram shown in Fig.~2
are the following:
\begin{eqnarray}
\label{ab}
G_{uu}^{\bar{\Lambda}} &=& G_{ud}^{\bar{\Lambda}} = G_{dd}^{\bar{\Lambda}} =
a_{\bar{\Lambda}}\cdot (1-z)^{\lambda+(\alpha_R-2\alpha_B) + 2(1-\alpha_B)+\Delta\alpha} \; , \\
G_{uu}^{\bar{\Xi}} &=& G_{ud}^{\bar{\Xi}} = G_{dd}^{\bar{\Xi}} = a_{\bar{\Xi}}
\cdot (1-z)^{\Delta 
\alpha}\cdot G_{uu}^{\bar{\Lambda}} \; , \\
G_{ds}^{\bar{\Lambda}} &=&
a_{\bar{\Lambda}}\cdot (1-z)^{\lambda+ (\alpha_R-2\alpha_B)+ 2(1-\alpha_B) + 
2\Delta\alpha}  \; , \\  
G_{ds}^{\bar{\Xi}} &=& \frac{a_{\bar{\Xi}}}{a_{\bar{\Lambda}}}\cdot (1-z)^{\Delta 
\alpha}\cdot G_{ds}^{\bar{\Lambda}} \;,
\end{eqnarray}

For the diquark fragmentation functions to baryon production, and as it was mentioned in the main
text, they have more complicated forms than the quark fragmentation functions, and they contain
two different contributions.
The first one corresponds to the central production of
one $B\bar{B}$ pair, and it is accounted for in Eq.~(4) by fragmentation
functions with the form:
\begin{eqnarray}
\label{t7}
G_{uu}^{\Lambda} &=& G_{ud}^{\Lambda} = G_{dd}^{\Lambda} = 
a_{\bar{\Lambda}}\cdot (1-z)^{\lambda-\alpha_R + 4(1-\alpha_B)+\Delta\alpha} \; , \\
%G_{uu}^{\bar{\Lambda}} &=& G_{ud}^{\bar{\Lambda}} = G_{dd}^{\bar{\Lambda}} =
%a_{\bar{\Lambda}}\cdot (1-z)^{\lambda+(\alpha_R-2\alpha_B) + 2(1-\alpha_B)+\Delta\alpha} \; ,
%%%\end{eqnarray}
%%%\newpage
%%%\begin{eqnarray}
G_{uu}^{\Sigma^+} &=& G_{dd}^{\Sigma^-} = G_{uu}^{\Sigma^-} =G_{dd}^{\Sigma^+}
= G_{ud}^{\Sigma^+} =G_{ud}^{\Sigma^-} = 
a_{\bar{\Sigma }^{\pm}}\cdot (1-z)^{\lambda-\alpha_R + 4(1-\alpha_B)+\Delta\alpha} \nonumber\; , \\
\\
G_{uu}^{\Xi^-} & =  & G_{ud}^{\Xi^-} =  G_{dd}^{\Xi^-} =
\frac{a_{\bar{\Xi}}}{a_{\bar{\Lambda}}}
\cdot (1-z)^{\Delta \alpha}\cdot G_{uu}^{\Lambda} \; , \\ 
%G_{uu}^{\bar{\Xi}} &=& G_{ud}^{\bar{\Xi}} = a_{\bar{\Xi}}\cdot (1-z)^{\Delta \alpha}G_{uu}^{\bar{\Lambda}} \; , \\
G_{uu}^{\Omega} &=& G_{ud}^{\Omega} =  G_{dd}^{\Omega} =
%G_{uu}^{\bar{\Omega}}= G_{ud}^{\bar{\Omega}} = 
a_{\bar{\Omega}}\cdot (1-z)^{\lambda-\alpha_R+4(1-\alpha_B)+3\Delta\alpha}\; , \\
%%%\end{eqnarray}
%%%\begin{eqnarray} 
G_{ds}^{\Lambda} &=& 
a_{\bar{\Lambda}}\cdot (1-z)^{\lambda-\alpha_R + 4(1-\alpha_B)+ 2\Delta\alpha} \; , \\
%G_{ds}^{\bar{\Lambda}} &=&
%a_{\bar{\Lambda}}\cdot (1-z)^{\lambda+ (\alpha_R-2\alpha_B)+ 2(1-\alpha_B) + 2\Delta\alpha} \cdot (1 + a_4z^{a_{4n}}) \; , \\
G_{ds}^{\Sigma^-} &=& G_{ds}^{\Sigma^+} = \frac{a_{\bar{\Sigma}}}{a_{\bar{\Lambda}}}
\cdot G_{ds}^{\Lambda} \; ,
\label{t9ds} \\ 
G_{ds}^{\Xi^-} &=& \frac{a_{\bar{\Xi}}}{a_{\bar{\Lambda}}}\cdot 
(1-z)^{\Delta \alpha}\cdot G_{ds}^{\Lambda} \;,
%\label{dsom1} \\
%G_{ds}^{\bar{\Xi}} &=& \frac{a_{\bar{\Xi}}}{a_{\bar{\Lambda}}}\cdot (1-z)^{\Delta \alpha}\cdot G_{ds}^{\bar{\Lambda}} \;,
\label{dsom} \\
G_{ds}^{\Omega} &=& \frac{a_{\bar{\Omega}}}{a_{\bar{\Xi}}}\cdot 
(1-z)^{\lambda-\alpha_R + 4(1-\alpha_B)+ 4\Delta\alpha}\;.
\end{eqnarray}

In these expressions we have used different parameterizations for the diquark fragmentation functions  to
antibaryon production and to central baryon production, since antibaryons are produced one cut kink
higher in the multiperipheral chain than the companion baryon (see Figure~\ref{figabbc}). The corresponding
expressions for proton and $\Lambda$ production in $pp$ collision were first given in ref.~\cite{KaPi}.

The second contribution in the diquark fragmentation functions to baryon production
comes from the direct fragmentation of the initial baryon into the secondary
one with conservation of $SJ$, shown in Figs.~\ref{f4}.
These contributions are determined by the following fragmentation functions:
\begin{eqnarray}
\label{ddL}
G_{dd}^{\Lambda} &=& a_N\cdot z^{\beta}\cdot
\Big[v_0^{\Lambda}\cdot\varepsilon\cdot (1-z)^2 + v_d^{\Lambda}\cdot 
z^{2 - \beta}\cdot (1-z) \nonumber \\
&+& v_{dd}^{\Lambda}\cdot z^{2.5 - \beta}\cdot (1-z)\Big]\cdot 
(1-z)^{\Delta \alpha} \; , \\
\label{dsL}
G_{ds}^{\Lambda} &=& a_N\cdot z^{\beta}\cdot
\Big[v_{0}^{\Lambda}\cdot\varepsilon\cdot (1-z)^{2+\Delta \alpha} + 
(v_{d}^{\Lambda}\cdot z^{2 - \beta}\cdot (1-z)^{1+\Delta \alpha} \nonumber \\ 
&+& v_{s}^{\Lambda}\cdot z^{2 - \beta+\Delta \alpha}\cdot (1-z))
+ v_{ds}^{\Lambda}\cdot z^{2.5-\beta}\cdot (1-\frac{z}{3})\Big] \; , \\
%\label{st14}
G_{uu}^{\Sigma^+} &=& G_{dd}^{\Sigma^-} = a_N\cdot z^{\beta}\cdot
\Big[v_{0}^{\Sigma^-}\cdot\varepsilon\cdot (1-z)^2 + v_{d}^{\Sigma^-}\cdot 
z^{2 - \beta}\cdot (1-z) \nonumber\\ 
&+& v_{dd}^{\Sigma^-}\cdot z^{2.5 - \beta} \Big]\cdot (1-z)^{\Delta \alpha} \;, \\
%\end{eqnarray}
%\begin{eqnarray}
G_{ud}^{\Sigma^+} &=& G_{ud}^{\Sigma^-} = a_N\cdot z^{\beta}\cdot
\Big[v_{0}^{\Sigma^-}\cdot\varepsilon\cdot (1-z)^2 \nonumber \\
&+& v_{d}^{\Sigma^-}\cdot z^{2 - \beta}\cdot
(1-z) \Big]\cdot (1-z)^{\Delta \alpha} \; , \\
G_{dd}^{\Sigma^+} &=& G_{uu}^{\Sigma^-} = a_N\cdot z^{\beta}\cdot
\Big[v_{0}^{\Sigma^-}\cdot\varepsilon\cdot(1-z)^2\Big]\cdot 
(1-z)^{\Delta \alpha} \; , \\
\label{dssigpl}
G_{ds}^{\Sigma^+} &=& a_N\cdot z^{\beta}\cdot
\Big[v_{0}^{\Sigma^+}\cdot\varepsilon\cdot (1-z)^{2+\Delta \alpha} \nonumber \\ 
&+& v_{s}^{\Sigma^+}\cdot z^{2 - \beta}\cdot(1-z)^{1+\Delta \alpha} +
v_{ds}\cdot z^{2.5-\beta}\cdot(1 - z)\Big] \; , \\
G_{ds}^{\Sigma^-} &=& a_N\cdot z^{\beta}\cdot\Big[v_{0}^{\Sigma^-}\cdot
\varepsilon\cdot (1-z)^{2+\Delta\alpha} 
+ (v_{d}^{\Sigma^-}\cdot z^{2-\beta}\cdot (1-z)^{1+\Delta\alpha} \nonumber \\
&+& v_{s}^{\Sigma^-}\cdot z^{2-\beta+\Delta\alpha}\cdot(1-z)^{1+\Delta \alpha}) 
+ v_{ds}\cdot z^{2.5-\beta} \Big] \;, \\
%%%\label{tdd14}
G_{dd}^{\Xi^-} &=& G_{ud}^{\Xi^-} = a_N\cdot z^{\beta}\cdot [v_0^{\Xi^-}\cdot\varepsilon 
\cdot (1-z)^2 + v_d^{\Xi^-}\cdot z^{2-\beta}\cdot (1-z)]\cdot (1-z)^{2\Delta\alpha} \;, 
\\
G_{ds}^{\Xi^-} &=& a_N\cdot z^{\beta}\cdot [v_0^{\Xi^-}\cdot \varepsilon 
\cdot (1-z)^{2+\Delta\alpha} + v_s^{\Xi^-}\cdot z^{2-\beta}\cdot(1-z)^{1+\Delta\alpha} 
\nonumber\\
&+&v_{ds}^{\Xi^-}\cdot z^{2.5-\beta} \Big]\cdot (1-z)^{\Delta\alpha} \;, \\
\label{GuuOm}
G_{uu}^{\Omega^-} &=& G_{ud}^{\Omega^-} = G_{dd}^{\Omega^-} =
a_N\cdot v_0^{\Omega^-}\cdot\varepsilon\cdot z^{\beta}\cdot
(1-z)^{2+3\Delta\alpha} \; , \\
\label{GdsOm}
G_{ds}^{\Omega} &=& a_N\cdot z^{\beta}\cdot[v_{0}\cdot\varepsilon \cdot 
(1-z)^{1+2\Delta \alpha} + v_{ds}\cdot z^{2-\beta}]\cdot {(1-z)^{1+2\Delta\alpha}}\; .
\end{eqnarray}

The third term in eqs.~(\ref{ddL}) and (\ref{dsL}) describes the contribution 
of the leading resonance $\Sigma^{*-}$ and its
subsequent decay into $\Lambda + \pi$ to the $dd$
and $ds$ fragmentation functions into $\Lambda$.
The factors $(1-z)$ in Eq.~(\ref{ddL}) and
$(1-\frac{z}{3})$ in Eq.~(\ref{dsL}) account for the suppression of this contribution.

%The values of the free parameters $a_1$ and $a_{1n}$ in the quark 
%fragmentation functions were determined in previous papers~\cite{ACKS,AMS} 
%and those of the free parameters $a_2$, $a_{2n}$, $a_3$, $a_{3n}$, $a_4$, 
%and $a_{4n}$ in the diquark fragmentation functions above have been obtained 
%here from comparison to experimental data,
%and they are given in Table~1.
%\vspace{-0.1cm}
%\begin{center}
%\begin{tabular}{|c|c|c|c|c|c|c|c|}\hline
% paremeter &valuep &  \\  \hline
%$a_1$ & $a_{1n}$ &$a_2$ & $a_{2n}$&$a_3$ & $a_{3n}$ &$a_4$ & $a_{4n}$ \\  
%\hline
%$12$ & $2$ & $2$ &$1$ & $3$ & $0.5$ & 20 & $2$  \\  \hline
%%\hline
%\end{tabular}
%\end{center}
%Table 1: The values obtained for the free parameters in the model by 
%comparison with experiment.

The different probabilities for the SJ without valence quarks, $v_0^B$, the SJ with
one valence quark, $v_q^B$, and the SJ with two valence quarks, $v_{qq}^B$, to go into
the secondary baryon $B$ were deduced through quark combinatorics~\cite{AS,CS}.
Here we assume that the strange quark suppression is common to the
three diagrams shown in Fig.~\ref{f4}, and thus, e.g. for the fragmentation of the SJ without
valence quarks into different baryons one gets:
\begin{equation}
p : n : \Lambda + \Sigma : \Xi^0 : \Xi^- : \Omega^- =
4L^3 : 4L^3 : 12L^2S : 3LS^2 : 3LS^2 : S^3 \;,
\label{relation}
\end{equation}
where the ratio $S/L$ determines the strange suppression factor, and 
$2L + S$ = 1. In the numerical calculations we have used $S/L = 0.32$.

Following the standard treatment in which the $\Sigma^0$ are included into $\Lambda$, and to
discriminate $\Lambda$ from the charged $\Sigma$, we use the
empirical rule: $\Sigma^+ + \Sigma^- = 0.6\cdot\Lambda$ \cite{CS}.

The values of $v_0^B$, $v_q^B$, and $v_{qq}^B$ used in Eq.~(9) are presented 
in Table~1.
\vspace{-0.1cm}
\begin{center}
%%%\vskip 12pt
\begin{tabular}{|c|c|c|c|c|c|c|c|c|}\hline
B & p & n & $\Lambda + \Sigma^0$ & $\Sigma^+$ & $\Sigma^-$ & $\Xi^0$ 
& $\Xi^-$ & $\Omega^-$  \\  \hline
$v_0$ & $4L^3$ & $4L^3$ & $7.5L^2S$ & $(9/4)L^2S$ & $(9/4)L^2S$ & $3LS^2$ 
& $3LS^2$  & $S^3$  \\  \hline
$v_u$ & $3L^2$ & $L^2$ & $(5/2)LS$ & $(3/2)LS$ & - & $S^2$ 
& - & -  \\  \hline
$v_d$ & $L^2$ & $3L^2$ & $(5/2)LS$ & - & $(3/2)LS$ & - 
& $S^2$  & -  \\  \hline
$v_s$ & - & - & $(5/2)L^2$ & $(3/4)L^2$ & $(3/4)L^2$ & $2LS$ 
& $2LS$  & $S^3$  \\  \hline
$v_{uu}$ & $2L$ & - & $(1/4)S$ & $(3/4)S$ & - & - 
& -  & -  \\  \hline
$v_{ud}$ & $L$ & $L$ & $S$ & - & - & - & - & -  \\  \hline
$v_{dd}$ & - & $2L$ & $(1/4)S$ & - & $(3/4)S$ & - & - & -  \\  \hline
$v_{us}$ & - & - & $(5/4)L$ & $(3/4)L$ & - & $S$ & - & -  \\  \hline
$v_{ds}$ & - & - & $(5/4)L$ & - & $(3/4)L$ & - & $S$ & -  \\  \hline
$v_{ss}$ & - & - & - & - & - & $L$ & $L$  & $S$  \\  \hline
\end{tabular}
%%%\vspace{-0.25cm}
\vspace{0.25cm}

\noindent
\hspace{-0.2cm}Table 1: The values of parameters $v_i^B$ in Eq.~(9) obtained 
from quark combinatorics.
%%%
%%%\noindent
%%%\hspace{-14.cm}\cite{AS,CS}.
\end{center}
%%%Table 2: The values of parameters $v_i^B$ in Eq.~(9) obtained from quark 
%combinatorics~\cite{AS,CS}.

Finally, we obtain from Eq.~(\ref{relation}) the relations among the values of the corresponding
parameters $a_{\bar{B}}$ in the fragmentation functions needed in Eq.~(4) for the production of
different $B\bar{B}$ pairs:
\begin{eqnarray}
a_{\bar{N}} : a_{\bar{\Lambda}} : a_{\bar{\Sigma}^{\pm}} : a_{\bar{\Xi}} : 
a_{\bar{\Omega}} =
1 &:& \sqrt{(15/8)\cdot (S/L)} : \sqrt{(9/16)\cdot(S/L)} : \sqrt{(3/4)\cdot(S/L)^2} \nonumber \\
&:& \sqrt{(3/4)\cdot(S/L)^2} : \sqrt{(1/4)\cdot(S/L)^3} \;.
\end{eqnarray}
% 15/8=5*((12/4)/1.6);9/16=(9/4)/4

%\newpage

\end{document}